\documentstyle[aps,prl,floats,graphicx,amssymb]{revtex}
\begin{document}
\draft
\twocolumn[\hsize\textwidth\columnwidth\hsize\csname
@twocolumnfalse\endcsname
\title{One-Dimensional Textures and Critical Velocity in  Superfluid
$^{3}$He-A}
\author{J. Kopu, R. H\"anninen, and E.~V. Thuneberg} 
\address{Low Temperature Laboratory, Helsinki University of 
Technology, 02150 Espoo, Finland}
\date{\today}
\maketitle
\begin{abstract} We study theoretically the stability of flow in
superfluid
$^{3}$He-A. The calculations are done using a one-dimensional model 
where the order parameter depends only on the coordinate in the
direction of the superfluid velocity ${\bf v}_{\rm s}$. We concentrate
on the case that the external magnetic field ${\bf H}$ is
perpendicular to ${\bf v}_{\rm s}$, where only few results are
available analytically.  We calculate the critical velocity $v_{c}$ at
which the superflow becomes unstable against the formation of
continuous vortices. The detailed dependence of $v_{c}$ on the
temperature and on the form of the underlying orbital texture
${\hat{\bf l}}({\bf r})$ is investigated. Both uniform and helical
textures of ${\hat{\bf l}}$ and two types of domain-wall structures
are studied. The results are partially in agreement with experiments
made in a rotating cylinder.

\end{abstract}
\pacs{PACS: 67.57.Pq}
\bigskip ] 

\section{Introduction}

The superflow of $^{3}$He-A differs markedly from the well known
superfluid $^{4}$He, where the decay of persistent currents is
prevented by topological constraints. The circulation of the
superfluid velocity
${\bf v}_{\rm s}$ on a closed contour is not quantized in $^{3}$He-A,
but depends on the field ${\hat{\bf l}}({\bf r})$ of the orbital
anisotropy vector. Thus $^{3}$He-A can respond to externally applied
flow by forming an inhomogeneous texture of 
${\hat{\bf l}}({\bf r})$. The rigidity of the
order parameter stabilizes the uniform bulk state (with
${\hat{\bf l}} ({\bf r})=$ constant) for small flow velocities in most
cases, depending on the magnitude and direction of the external magnetic
field ${\bf H}$.  With increasing velocity this configuration becomes
unstable against textural inhomogeneities that finally lead to the
formation of ``continuous'' vortices. In several cases a helical
texture of ${\hat{\bf l}}({\bf r})$ is stabilized at intermediate
velocities. 

The flow properties of bulk
$^{3}$He-A were studied in many theoretical papers
which peak in the late 1970's  
\cite{Bhatta,CL,HSol,LinLiu,Maki,JLTP,Fetter,FW,Fetter2,DH,DL,1988}.
The majority of that work studied the
case where the superfluid velocity and the magnetic field are
parallel.  Various methods were used to generate the flow
experimentally  (torsional oscillator, thermal gradient, a piston, etc.
see Refs.\
\cite{HHhydro,Bozler} for reviews). Although some of the predicted
features were seen in the experiments, no satisfactory agreement
between theory and experiment was found. 

More recently, a rotating cryostat has been used to generate superflow
\cite{Ruutu}. This method gives a well defined dc superfluid velocity
under very steady conditions. These experiments motivated our
theoretical studies. Firstly, it was necessary to study the case where
${\bf v}_{\rm s}$ is perpendicular to ${\bf H}$. The helical texture
in this case was previously considered only in one paper
\cite{LinLiu}. Secondly, the calculation has to be generalized to
temperatures substantially below the superfluid transition temperature
$T_{\rm c}$. Thirdly, we consider two different types of ``solitons''.
These are domain-wall like structures whose effect on the flow
properties were previously considered in a few cases
\cite{HSol,Maki,DH,DL,1988}.  Here we present detailed calculations
of the critical velocities
$v_{\rm c}$ for the appearance of helical textures and of vortices in
different cases with ${\bf v}_{\rm s}\perp{\bf H}$. The comparison of
the results with experimental measurements has been given before
\cite{Ruutu}.

The hydrodynamic free energy describing the current-carrying states of
$^{3}$He-A is recalled in Sec.\
\ref{energy}. In Sec.\ \ref{calcul} we study the general features of
our one-dimensional model and its numerical solution. Sec.\
\ref{Helical} discusses the instabilities of uniform and helical
textures. In Sec.\ \ref{soliton} the critical velocity associated with
two initially inhomogeneous textures, a dipole-locked and a
dipole-unlocked soliton, is investigated. The comparison to
experiments is briefly discussed in Sec.\
\ref{conc}.

\section{Hydrostatic theory}
\label{energy}

The order parameter of $^{3}$He-A is fully specified by defining an
orthonormal triad $\{ {\hat{\bf m}}$, ${\hat{\bf n}}$, ${\hat{\bf l}}
\}$ and a unit vector ${\hat{\bf d}}$. The triad describes the orbital
part and ${\hat{\bf d}}$ the spin part of the tensor order parameter
\cite{Voll}
\begin{equation} A_{\mu j} = \Delta \hat{d}_{\mu} (\hat{m}_{j} + {\rm
i}\hat{n}_{j}).
\label{e.op}\end{equation} Here $\Delta$ is a (temperature-dependent)
constant and ${\hat{\bf l}}\equiv {\hat{\bf m}}\times{\hat{\bf n}}$.
All the unit vectors may vary as functions of location ${\bf r}$.  A
change in the total phase $\Phi$ of the order parameter (\ref{e.op})
is given by
\begin{equation}
\delta\Phi =   {1 \over 2}\sum_j\left[\hat m_j \delta \hat n_j - \hat
n_j \delta \hat m_j \right] = 
\sum_j\hat m_j \delta \hat n_j.
\label{e.phase}\end{equation} This allows for the definition for the
superfluid velocity as
\begin{equation} {\bf v}_{\rm s}=  {\hbar\over 2m}\sum_j\hat
m_j\bbox{\nabla}\hat n_j, 
\label{supvel}
\end{equation} where $m$ in the prefactor is the mass of a $^{3}$He
atom.  The equilibrium properties of the superfluid are determined by
the order-parameter configuration that minimizes the total free
energy. In the hydrodynamic approximation it has the form
\begin{equation} F = 
\int {\rm d}^{3}{\bf r} \hspace{1mm} f =
\int {\rm d}^{3}{\bf r} \hspace{1mm} ( f_{\rm gr} + f_{\rm d} + f_{\rm
h} ).
\label{ftot}
\end{equation} The gradient energy density can be written as 

\begin{eqnarray} f_{\rm gr}&=&\textstyle{1\over 2}
\rho_\perp{\bf v}_{\rm s}^2+\textstyle{1\over 2}
(\rho_\parallel-\rho_\perp)(\hat{\bf l}\cdot{\bf v}_{\rm s})^2
\nonumber \\&\hbox{}+&C{\bf v}_{\rm s}\cdot\nabla\times\hat{\bf l} -
C_0(\hat{\bf l}\cdot{\bf v}_{\rm s}) (\hat{\bf
l}\cdot\nabla\times\hat{\bf l})\nonumber \\&\hbox{}+&
\textstyle{1\over 2}K_{\rm s}(\nabla\cdot\hat{\bf l})^2 +
\textstyle{1\over 2}K_{\rm t}(\hat{\bf l}\cdot\nabla\times\hat{\bf
l})^2+\textstyle{1\over 2}K_{\rm b}\vert\hat{\bf
l}\times(\nabla\times\hat{\bf l})\vert^2 \nonumber \\ &\hbox{}+&
\textstyle{1\over 2}K_5 \vert(\hat{\bf l}\cdot\nabla)\hat{\bf
d}\vert^2+ \textstyle{1\over 2}K_6\sum_{ij}[(\hat{\bf
l}\times\nabla)_i\hat{\bf d}_j]^2. 
\label{fgrad}
\end{eqnarray} The first four terms give the kinetic energy of the
anisotropic superfluid.
The $C$ and $C_0$ terms give coupling between flow and an
inhomogeneous ${\hat{\bf l}}$ field. The remaining five terms are the
bending energy densities for ${\hat{\bf l}}$ and ${\hat{\bf d}}$.

The energy density of dipole-dipole interaction is
\begin{equation} f_{\rm d} = -\textstyle{1\over 2}g_{\rm d}(\hat{\bf
d}\cdot\hat{\bf l})^2.
\label{fd}
\end{equation} Comparing this to the kinetic energy (\ref{fgrad})
defines the dipole velocity $v_{\rm d}= \sqrt {g_{\rm
d}/\rho_{\parallel}} \sim 1$ mm/s and the dipole length 
$\xi_{\rm d}=\hbar/2mv_{\rm d} \sim 10\ \mu$m. In addition, the energy
density in the presence of an external magnetic field ${\bf H}$ is
\begin{equation} f_{\rm h} = \textstyle{1\over 2}g_{\rm h}(\hat{\bf
d}\cdot{\bf H})^2.
\label{fh}
\end{equation} It is customary to define the dipole field
$H_{\rm d} =\sqrt{g_{\rm d}/g_{\rm h}}\sim$ 2 mT by comparing
(\ref{fd}) and (\ref{fh}).  

All the coefficients $\rho_\perp$, $\rho_\parallel$, $C$, $C_0$,
$K_{\rm s}$, $K_{\rm t}$, $K_{\rm b}$, $K_5$, $K_6$, $g_{\rm d}$, and
$g_{\rm h}$  are positive and depend on the temperature $T$ and the
pressure $p$. Their values are determined as explained in Ref.
\cite{Parts}. In particular, the coefficients are calculated using
consistently the weak-coupling approximation. In reduced units of
$v_{\rm d}$, $\xi_{\rm d}$ and $H_{\rm d}$ our results are independent
of $g_{\rm d}$ and $g_{\rm h}$, but they are needed for  comparison to
experiments \cite{Ruutu}. For 
$g_{\rm d}$ we write
\begin{equation} g_{\rm d}(T,p) = 4g_{\rm D}^0 (p)
\hspace{1mm} \Delta_{\rm A}^2 (T,p),
\end{equation} where $\Delta_{\rm A}$ is the maximum energy gap in the
weak-coupling approximation. All calculations are done at the melting
pressure where
$g_{\rm D}^0=5.9\cdot 10^{44}\ {\rm J^{-1} m^{-3}}$ \cite{hydrostatic}.
We wish to emphasize that the calculations contain no adjustable
parameters.

The hydrostatic theory gives a good description of the superflow in
$^3$He-A over most of the temperature region $0<T<T_{\rm c}$. It becomes
invalid in a small region $T_{\rm c}-T\lesssim 10^{-6}T_{\rm c}$ around
$T_{\rm c}$, where the Ginzburg-Landau critical velocity 
$\sim{\hbar\over 2m}\xi_{\rm GL}^{-1}$ is smaller than $v_{\rm d}$.
We also  limit to such low fields that the deformation of the A  phase
order parameter (\ref{e.op}) towards the A$_1$ phase can be neglected.

\section{One-dimensional calculation}
\label{calcul}

We study flow in bulk $^3$He-A far from any walls. The main assumption
in the present work is that the order parameter (\ref{e.op}) depends
only on one spatial coordinate
$x$. It follows from the definition (\ref{supvel}) that ${\bf v}_{\rm
s}$ is always parallel to the $x$ axis, ${\bf v}_{\rm
s}\parallel\hat{\bf x}$.   In addition to the homogeneous state, the
1D model allows us to calculate the structure of helical textures and
the deformation of solitons in a flow that is perpendicular to the
soliton wall. Moreover, we also can determine the stability limits of
such textures against 1D perturbations. We will argue in section
\ref{conc} that the local stability of the states we consider is indeed
determined by such perturbations.

The cases of $H=0$
\cite{Bhatta,CL,Fetter,DH,1988} and
${\bf H}\parallel {\hat{\bf x}}$
\cite{HSol,LinLiu,Maki,JLTP,FW,Fetter2,DL} have been studied
extensively in the literature . Here we study the more complicated case
${\bf H} \perp {\hat{\bf x}}$
\cite{LinLiu,Maki}. We study in particular the
high field limit $H
\gg H_{\rm d}$, where ${\hat{\bf d}} \perp {\bf H}$ everywhere.

The order parameter of
$^{3}$He-A can be parametrized by introducing three Euler angles
$\alpha$, $\beta$ and $\gamma$ for the orbital triad
$\{ {\hat{\bf m}}$, ${\hat{\bf n}}$, ${\hat{\bf l}} \}$ and
polar and azimuthal angles $\psi$ and $\phi$ for ${\hat{\bf d}}$.
In this
representation one has to beware the unphysical singularities at the
poles $\sin\beta=0$ or $\sin\psi=0$. For this reason we choose the
polar axis ${\hat{\bf z}}$ of the fixed coordinate system 
perpendicular to the direction of ${\bf v}_{\rm s}$ (and parallel to
the external magnetic field). This results in a more complicated form
of the energy functional but it enables us to avoid the singularities
in all stationary states. With these definitions
\begin{eqnarray} {\hat {\bf l}} &=&\sin\beta \cos\alpha \hspace{1mm}
{\hat {\bf x}} + \sin\beta \sin\alpha \hspace{1mm} {\hat {\bf
y}}+\cos\beta 
\hspace{1mm} {\hat {\bf z}} \label{lpara}  \\ {\hat {\bf d}}& =&
\sin\psi \cos\phi \hspace{1mm} {\hat {\bf x}} + \sin\psi \sin\phi
\hspace{1mm} {\hat {\bf y}} + \cos\psi 
\hspace{1mm} {\hat {\bf z}}   \label{dpara}\\ {\bf v}_{\rm
s}&=&-{\hbar \over 2m} (\frac{{\rm d}\gamma}{{\rm d}x} + \cos \beta
\hspace{1mm} \frac{{\rm d}\alpha}{{\rm d}x}) \hspace{1mm} {\hat {\bf
x}}.  
\label{para}
\end{eqnarray} The form of the unknown functions $\alpha(x)$,
$\beta(x)$,
$\gamma(x)$, $\psi(x)$ and $\phi(x)$ in equilibrium corresponds to the
minimum of the free energy functional (\ref{ftot}) where
\begin{equation}
\frac{\delta f}{\delta \alpha} \equiv \frac{\partial f} {\partial
\alpha} - \frac{\rm d}{{\rm d}x} \left[ 
\frac{\partial f}{\partial ({\rm d}\alpha/{\rm d}x)} \right] = 0,
\label{euler}
\end{equation} and similarly for other angles.

We point out two analytic observations that are useful in testing the
convergence and the accuracy of the numerical solution. Firstly, the
superfluid velocity ${\bf v}_{\rm s}$ (\ref{para}), and therefore also
the total free energy (\ref{ftot}), do not depend explicitly on the
angle
$\gamma$ but only on its derivative. The corresponding Euler equation
(\ref{euler}) reduces to a conservation law
\begin{equation} {\partial f \over \partial  ({\rm d}\gamma / {\rm
d}x)} \equiv -p = {\rm const}.
\label{curr}
\end{equation} The quantity $({2m/\hbar})p$ is the $x$ component of
the supercurrent density
\begin{eqnarray}
\nonumber {\bf j}_{\rm s}
 &=& \rho_\perp{\bf v}_{\rm s}+ (\rho_\parallel-\rho_\perp)\hat{\bf
l}(\hat{\bf l}\cdot{\bf v}_{\rm s}) \\ &+&C \nabla\times\hat{\bf
l}-C_0 \hat{\bf l} (\hat{\bf l}\cdot\nabla\times\hat{\bf l}).
\end{eqnarray} Another conserved quantity in the problem is the one
corresponding to the fact that the free energy (\ref{ftot}) does not
depend explicitly on
$x$ either. The invariant related to this is analogous to the
Hamiltonian of classical mechanics and can be brought to the form
\begin{equation} f_{\rm gr} - f_{\rm d} - f_{\rm h} = {\rm const.}
\label{hamil}
\end{equation}

We wish to study a one-dimensional interval of length $L \gg \xi_{\rm
d}$. A flow through the system is achieved by keeping a fixed phase
difference $\Delta\Phi \equiv \Phi(L/2)-\Phi(-L/2)$ between the
endpoints of the line. The constancy of $\Delta\Phi$ as a function of
time $t$ is enforced by imposing the boundary condition 
\begin{eqnarray} {{\rm d}\Delta\Phi \over {\rm
d}t}&\equiv&-\left({{\rm d}\gamma \over {\rm d}t}+
\cos\beta{{\rm d}\alpha \over {\rm d}t}\right)_{x={L \over 2}}+ 
\left({{\rm d}\gamma \over {\rm d}t}+
\cos\beta{{\rm d}\alpha \over {\rm d}t}\right)_{x=-{L \over
2}}\nonumber\\ &=&0.
\label{BC}
\end{eqnarray} It is, however, more advantageous to express the
results in terms of a driving velocity defined by  $v_{\rm
n}=(\hbar/2m)
\Delta\Phi/L$. This is a more convenient quantity than $\Delta\Phi$
because all our results are independent of $L$ when expressed in
terms of $v_{\rm n}$. $v_{\rm n}$ could be identified as the velocity
of the normal component that drives the superfluid component of the
liquid. However, $v_{\rm n}$ should be considered as a scalar
parameter since all vector quantities in this paper (like ${\bf
v}_{\rm s}$ and ${\bf j}_{\rm s}$) are given in the frame where the
normal fluid is at rest, ${\bf v}_{\rm n}\equiv 0$. (See Ref.\
\cite{Karimaki} for a general formulation with
${\bf v}_{\rm n}\not= 0$.) 

In general case we have to use numerical methods to determine the
equilibrium state of the system. The order parameter is taken to be
defined at $N$ equally spaced discrete points on the line. The
discretization length $\Delta x \equiv L/N$ is chosen much smaller
than the dipole length, usually
$\Delta x \lesssim 0.1 \xi_{\rm d}$. As a first step we have to choose
some initial configuration for the five angles. A given $\Delta\Phi$ or
$v_{\rm n}$ is implemented by taking an initial guess $\gamma(x)=
-{2m\over\hbar}v_{\rm n} x$. The angle functions are then iterated
numerically towards the equilibrium state for a given $v_{\rm n}$ using
the following diffusion-like equations
\begin{eqnarray}
\nonumber
\mu_{1} \sin^{2} \beta \hspace{1mm} 
\frac{\partial \alpha}{\partial t} &=& -\frac{\delta f}{\delta \alpha}
\\
\nonumber
\mu_{1} \frac{\partial \beta}{\partial t} &=& -\frac{\delta f}{\delta
\beta} \\
\nonumber
\mu_{2} \frac{\partial \gamma}{\partial t} &=& -\frac{\delta f}{\delta
\gamma} \\
\nonumber
\mu_{3} \sin^{2} \psi \hspace{1mm} 
\frac{\partial \phi}{\partial t} &=& -\frac{\delta f}{\delta \phi}
\nonumber \\
\mu_{3} \frac{\partial \psi}{\partial t} &=& -\frac{\delta f}{\delta
\psi} 
\label{dyna}
\end{eqnarray} together with the boundary condition (\ref{BC}). The
simplest discretized expressions have been used in representing the
derivatives in equations (\ref{dyna}). Since we do not attempt to
describe the true time evolution of the textures, the viscosity
constants $\mu_1$, $\mu_2$ and $\mu_3$ can be chosen according to
numerical convenience.

\section{Uniform and Helical textures}
\label{Helical}

\begin{figure}[tbh] 
\begin{center}\leavevmode
\includegraphics[width=0.75\linewidth]{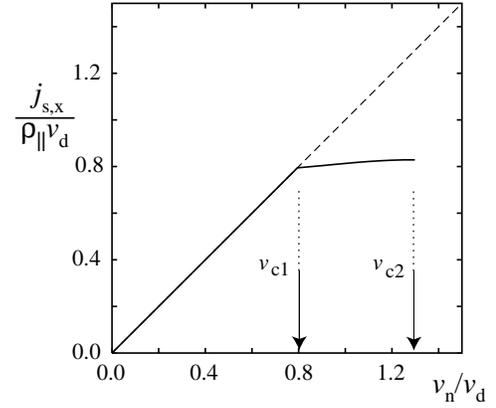} \bigskip
\caption[virrat]{ The supercurrent $j_{{\rm s},x}$  as a function of
driving velocity $v_{\rm n}$ at $T=0.6T_{\rm c}$ and
$H\gg H_{\rm d}$.  The diagonal line corresponds to the current in the
uniform texture, 
$j_{\rm s}=\rho_\parallel v_{\rm n}$. The solid line between $v_{\rm
c1}$ and $v_{\rm c2}$ corresponds to helical texture at the optimal wave
vector
$q_{\rm opt}$. }\label{virrat}\end{center}\end{figure}

The simplest texture has constant ${\hat{\bf l}}\parallel{\hat{\bf
d}}\parallel\hat{\bf x}$. This uniform state minimizes both the
dipole-dipole energy (\ref{fd}) and the field energy (\ref{fh}) for
${\bf H} \perp {\hat{\bf x}}$. It also corresponds to the
minimum of the first two terms in the gradient energy (\ref{fgrad})
because $\rho_\parallel < \rho_\perp$.
The current in this state is linear in 
$v_{\rm n}$, ${\bf j}_{\rm s}=\rho_\parallel v_{\rm n}\hat{\bf x}$, as
illustrated in Fig.\ \ref{virrat}.

\begin{figure}[tbh]
\begin{center}\leavevmode
\includegraphics[width=0.9\linewidth]{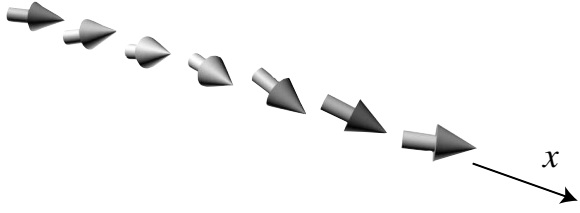}
\bigskip
\caption[helic]{ Variation of ${\hat{\bf l}}$ in a helical texture in
one spatial dimension (${\hat{\bf x}}$ is in the direction of flow).
One wavelength $\lambda$ of the helix is shown.
}\label{helic}\end{center}\end{figure}

The uniform state is stable at small velocities (if $H\not= 0$). We
call the stability limit of the uniform texture as the first critical
velocity
$v_{\rm c1}$. At $v_{\rm n}=v_{\rm c1}$ the uniform state becomes
unstable against a helical deformation where
${\hat{\bf l}}$ winds around the direction of the flow, see 
Fig.\ \ref{helic}.
At this point the energy cost in forming an inhomogeneous texture is
compensated by reductions in other energy terms. In particular, the
superfluid velocity $v_{\rm s}$ (\ref{para}) is lowered for a given
$v_{\rm n}$ and there is a negative contribution from the energy term
with the coefficient
$C_{0}$ (\ref{fgrad}) \cite{Bhatta}. 

The instability point  can be studied by expanding the energy
(\ref{ftot}) around the uniform solution. This calculation was done by
Lin-Liu et al in the Ginzburg-Landau region \cite{LinLiu}. We
generalize this calculation to all temperatures.  We define $\gamma$
as above but otherwise use a parametrization that is different from
(\ref{lpara}-\ref{para}):
\begin{eqnarray} {\hat {\bf l}} &=&[1-\textstyle{1\over 2}(l_y^2+l_z^2)
-{1\over 8}(l_y^2+l_z^2)^2] {\hat {\bf x}} +l_y{\hat {\bf y}} +
l_z{\hat {\bf z}}\label{lpara2}  \\ {\hat {\bf d}}& =&
[1-\textstyle{1\over 2}(d_y^2+d_z^2) -{1\over 8}(d_y^2+d_z^2)^2]  
 {\hat {\bf x}}   +d_y{\hat {\bf y}} + d_z{\hat {\bf z}}\\ {\bf
v}_{\rm s}&=&{\textstyle{\hbar \over 2m}}\{-\frac{{\rm d}\gamma}{{\rm
d}x}+{\textstyle{1\over 2}}(l_y\frac{{\rm d}l_z}{{\rm d}x}
-l_z\frac{{\rm d}l_y}{{\rm d}x})\textstyle[1+{1\over 4}(l_y^2+l_z^2)
]\} {\hat {\bf x}}.  
\label{para2}
\end{eqnarray} These equations are valid up to fourth order in $l_y$,
$l_z$, $d_y$, and $d_z$. We eliminate $\gamma(x)$ in favor of the
current $p$ (\ref{curr}) by defining a new free energy
$G=F+\int{\rm d}x({\rm d}\gamma/{\rm d}x)p$. Substitution of
(\ref{lpara2}-\ref{para2}) into $G$ and expansion to second order
gives linear Euler-Lagrange equations. These have the solution
\begin{eqnarray} l_y(x)&=&us\sin(qx)  \\ l_z(x)&=&(u/s)\cos(qx)\\
d_y(x)&=&u\delta_1s\sin(qx)  \\ d_z(x)&=&(u\delta_2/s)\cos(qx)
\label{para3}
\end{eqnarray} where $q$ is the wave vector of the helix. The 
magnetic field in the transverse $z$ direction introduces ``easy'' and
``hard'' directions for the amplitudes, and thus the helix has an
elliptically distorted form.  When $G$ is minimized with respect to
$s$, $\delta_1$, and $\delta_2$ we find
\begin{eqnarray}
\delta_1 &=& (K_5 q^2+1)^{-1} \label{e.shortper} \\
\delta_2 &=& (K_5 q^2+H^2+1)^{-1}  \\ s^2&=&K_2/K_1  \\ K_i^2 &=& 1 +
(\rho_\perp-1)p^2 + K_{\rm b} q^2 - \delta_i   \ .
\end{eqnarray} For the free energy we find the expansion
\begin{eqnarray} G= G_0+\textstyle{1\over
2}Au^2+\textstyle{1\over 4}Bu^4
\label{Gexp}
\end{eqnarray} where 
\begin{eqnarray} &G_0 &= -\textstyle{1\over 2}(1+p^2)  \\  &A &= K_1
K_2 - (2C_0 +1) qp  \label{acoeffiecient}\\ &B &= \left[
(2\rho_\perp-1)C_0 - {\textstyle\frac{1}{4}} \right]
\left(\frac{K_2}{K_1} + \frac{K_1}{K_2}\right) qp  +
{\textstyle\frac{1}{4}}\Bigg\{
\nonumber \\  &&+  \frac{K_2^2}{K_1^2} \big[ 3\delta_1(1-\delta_1)^2 +
(K_{\rm s} +(K_6-K_5)\delta_1^2
\nonumber \\ &&+ K_5\delta_1^4) q^2 - 3(\rho_\perp-1)^2 p^2
\big]\nonumber \\ &&+  \frac{K_2^2}{K_1^2} \big[
3\delta_2(1-\delta_2)^2 + (K_{\rm s} + (K_6-K_5)\delta_2^2 
\nonumber \\&&+ K_5\delta_2^4) q^2 - 3(\rho_\perp-1)^2 p^2
\big]\nonumber \\ &&+  \big[ 8K_{\rm t} - 2K_{\rm s} - 8C_0^2 -
8K_{\rm b} \nonumber
\\&&+ 3(K_6-K_5)(\delta_1^2+\delta_2^2)  -2K_5 \delta_1^2 \delta_2^2
\big] q^2 \nonumber \\  &&+  
(1-\delta_1)(1-\delta_2)(\delta_1+\delta_2) -2(\rho_\perp-1)^2 p^2 
\Bigg\}
\ . 
\label{e.abper}\end{eqnarray} For simplicity, we have used units
where $g_{\rm d} = g_{\rm h} = \rho_{\parallel} =
\frac{\hbar}{2m} = 1$ in Eqs.\ (\ref{e.shortper})-(\ref{e.abper}).
Near the superfluid transition temperature these results reduce to
those by Lin-Liu et al \cite{LinLiu}. 

\begin{figure}[tb]
\begin{center}\leavevmode
\includegraphics[width=0.75\linewidth]{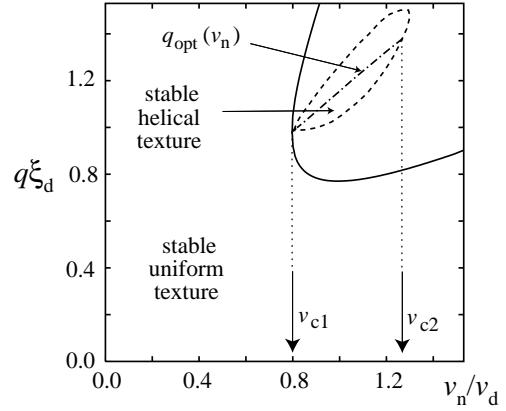}
\bigskip
\caption[qkuva]{  The stability of different states in the $v_{\rm n}
- q$ plane  at $T = 0.6 T_{\rm c}$ and for 
$H\gg H_{\rm d}$.  The solid line marks the instability of the uniform
texture according to $A(v_{\rm n},q)=0$ (\ref{acoeffiecient}). The
helical texture is stable inside the region spanned by the dashed
line. The dash-dotted line shows the optimum wave vector
$q_{\rm opt}$ as a function of the driving velocity $v_{\rm n}$.
}\label{qkuva}\end{center}\end{figure}

\begin{figure}[tb]
\begin{center}\leavevmode
\includegraphics[width=0.95\linewidth]{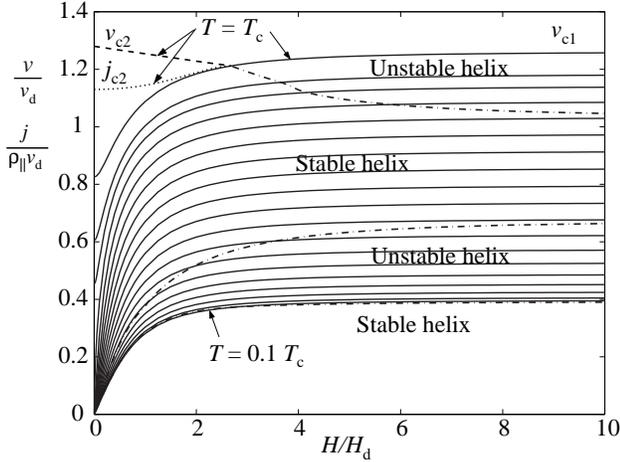}
\bigskip
\caption{  The critical velocity $v_{\rm c1}$ of the uniform texture
(solid lines) as a function of the magnetic field $H$. The different
curves correspond to different temperatures with intervals of
$0.05T_{\rm c}$. The regions of stability and instability of a
small-angle helix are separated by dash-dotted lines.  The dashed line
denotes $v_{\rm c2}$ at $T=T_{\rm c}$.  The corresponding critical
current 
$j_{\rm c2}$ in units of $\rho_\parallel v_{\rm d}$ is indicated by
dotted line.
 }\label{risto}\end{center}\end{figure}

The uniform texture is stable if the coefficient $A$ is positive. The
line where $A$ vanishes in the $v_{\rm n}$--$q$ plane is shown by
solid line in Fig.\ \ref{qkuva}.
In a long interval $L\gg\xi_{\rm d}$ the value of the wave vector $q$
is not limited. This means that the uniform texture is stable only
below the critical velocity
$v_{\rm c1}$ defined by the conditions $A={\partial A/\partial q}=0$.
The velocity
$v_{\rm c1}$ is plotted as a function of magnetic field and
temperature in Fig.\ \ref{risto}.
Note that $v_{\rm c1}$ vanishes in zero field for temperatures
$T\lesssim 0.85T_{\rm c}$.

The stability of small-angle helical textures is determined by the
coefficient $B$. The helix is stable if $B>0$ and unstable if $B<0$.
The stability as a function of $T$ and $H$ is indicated in Fig.\
\ref{risto}.

Helical textures with general opening angles were studied numerically.
The periodicity of the helix allows the numerical calculations to be
limited to a single wavelength
$\lambda=2\pi/q$ using periodic boundary conditions for ${\hat{\bf
l}}$ and
${\hat{\bf d}}$. In fact, making use of all the symmetries even a
quarter of $\lambda$ would be sufficient. We then minimize the free
energy (\ref{ftot}) with respect to the five angle fields in
(\ref{lpara})-(\ref{para}). This gives the energy as a function of
$v_{\rm n}$ and $q$: $F(v_{\rm n},q)$.  

For each value of $v_{\rm n}$ we determine the optimum wave vector
$q_{\rm opt}$ at which the energy of the helix is minimized. This
process is simplified by the fact that the derivative of $F$ with
respect to $q$ can be obtained using the formula
\begin{equation} {\partial F \over \partial q} = {1 \over q}\left(
2F_{\rm gr}- v_{\rm n} J_{{\rm s},x} \right),
\label{qder}
\end{equation} 
 with $F_{\rm gr}$ and $J_{{\rm s},x}$ defined to be the corresponding
densities $f_{\rm gr}$ and $j_{{\rm s},x}$ integrated over the
one-dimensional interval. At the optimum value $q=q_{\rm opt}(v_{\rm
n})$ the derivative given by (\ref{qder}) vanishes.  The dependence of
$q_{\rm opt}$ on
$v_{\rm n}$  is illustrated    in Fig.\ \ref{qkuva}. 

\begin{figure}[tbh]
\begin{center}\leavevmode
\includegraphics[width=0.93\linewidth]{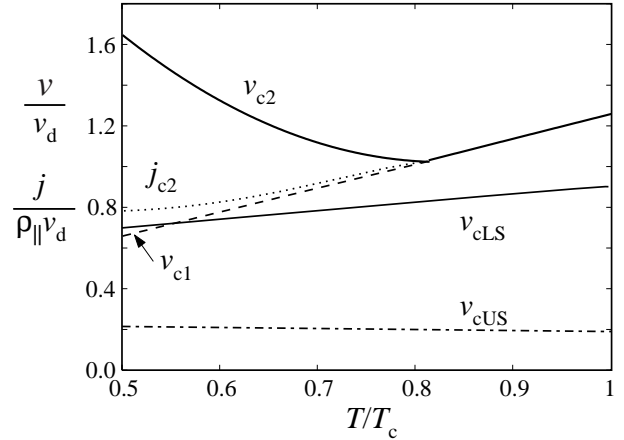}
\bigskip
\caption[helical]{ The critical velocity as a function of temperature
$T$ for $H\gg H_{\rm d}$. The different curves correspond to
instabilities of the uniform texture $v_{\rm c1}$, the helical texture
$v_{\rm c2}$, the locked soliton $v_{\rm cLS}$, and the unlocked
soliton $v_{\rm cUS}$. Helical textures are stable in the region
between $v_{\rm c1}$ and
$v_{\rm c2}$. The dotted line is 
$j_{\rm c2}$ in units of $\rho_\parallel v_{\rm d}$.
}\label{helical}\end{center}\end{figure}

The stability of helical textures is determined by the eigenvalues of
the Hessian matrix of $F(v_{\rm n},q)$:
\begin{displaymath} {\cal H} = \left( 
\begin{large}
\begin{array}{cc}
\frac{\partial^{2}F}{\partial {v_{\rm n}}^{2}} &
\frac{\partial^{2}F}{\partial v_{\rm n} \partial q} \\ & \\
\frac{\partial^{2}F}{\partial v_{\rm n} \partial q} &
\frac{\partial^{2}F}{\partial q^{2}}
\end{array} 
\end{large} \right).
\end{displaymath} A texture is stable if both the eigenvalues of the
Hessian matrix are positive, and unstable otherwise. The stability
region is indicated by a dashed line in Fig.\
\ref{qkuva}. The velocity where the state at optimal wave vector
$q_{\rm opt}$ becomes unstable is defined as the second critical
velocity, $v_{\rm c2}$. In helical texture the current increases much
slower with $v_{\rm n}$ than in the uniform state (Fig.\ 
\ref{virrat}). Therefore, the critical current $j_{\rm c2}$ is
substantially smaller than $\rho_\parallel v_{\rm c2}$. At
$H=0$ and $T\approx T_{\rm c}$ we find that $v_{\rm c2}\approx
1.28v_{\rm d}$ whereas
$j_{\rm c2}\approx 1.13\rho_\parallel v_{\rm d}$ \cite{JLTP} (Fig.\
\ref{risto}). In general we find essentially no field dependence of
$v_{\rm c2}$, in contrast to $v_{\rm c1}$ which vanishes at $H=0$ when
$T\lesssim 0.85T_{\rm c}$. The temperature dependencies of
$v_{c1}$, $v_{c2}$ and $j_{c2}$ in the high field limit are presented
in Fig.\ \ref{helical}.
At high temperatures $T>0.8T_{\rm c}$  there is no stable helical
texture and thus
$v_{\rm c1}$ and $v_{\rm c2}$ coincide. At lower temperatures $v_{\rm
c2}$ is seen to grow distinctly above $v_{\rm c1}$.   Below $0.5T_{\rm
c}$ helical textures again become unstable (Fig.\
\ref{risto}).

The numerical calculation of the Hessian matrix is simplified by the
fact that both first derivatives (\ref{curr}) and (\ref{qder}) are
easily available. Thus the calculation of the Hessian at point
$(v_{\rm n},q)$ requires texture minimizations only at three points:
$(v_{\rm n},q)$,
$(v_{\rm n},q+\Delta q)$ and $(v_{\rm n}+\Delta v_{\rm n},q)$. In
order to minimize errors the number $N$ of discretization points within
a wave length $\lambda$ was kept constant  and the discretization
length
$\Delta x=\lambda/N$ was varied instead.

\begin{figure}[bt]
\begin{center}\leavevmode
\includegraphics[width=\linewidth]{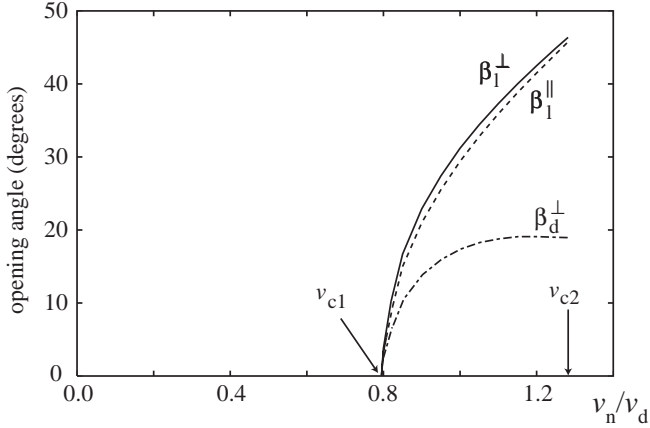}
\bigskip
\caption[kulmat]{  The opening angles of ${\hat{\bf l}}$ in the plane
$\perp {\bf H}$ ($\beta_{\rm l}^{\perp}$, solid line) and $\parallel
{\bf H}$ ($\beta_{\rm l}^{\parallel}$, dashed line) and the opening
angle of
${\hat{\bf d}}$ in the plane $\perp {\bf H}$ ($\beta_{\rm d}^{\perp}$,
dash-dotted line) as functions of driving velocity $v_{\rm n}$ with
$q=q_{\rm opt}$, $T=0.6T_{\rm c}$ and 
$H\gg H_{\rm d}$.  }\label{kulmat}\end{center}\end{figure}

For all helices the opening angle grows continuously from zero with
increasing $v_{\rm n}$, see Fig.\ 
\ref{kulmat}.
The largest stable values
for the opening angle found in the simulations were
$\sim 60$ degrees. We have made numerical simulations in an interval
containing several wave lengths of the helix, $L\gg\lambda$. When the
limit of stability is exceeded, it seems that the number of windings
of the helix changes if a stable texture is possible at a given
$v_{\rm n}$. Otherwise, the instability seems to lead to the growth of
the opening angle. Most likely this leads to formation of continuous
vortex structures
\cite{Karimaki}. However, we were  not able to follow this process
beyond $90^\circ$ opening angles because of the singularity in the
coordinate system (\ref{lpara})-(\ref{para}).

\section{Soliton textures}
\label{soliton}

\begin{figure}[tbh]
\begin{center}\leavevmode
\includegraphics[width=0.9\linewidth]{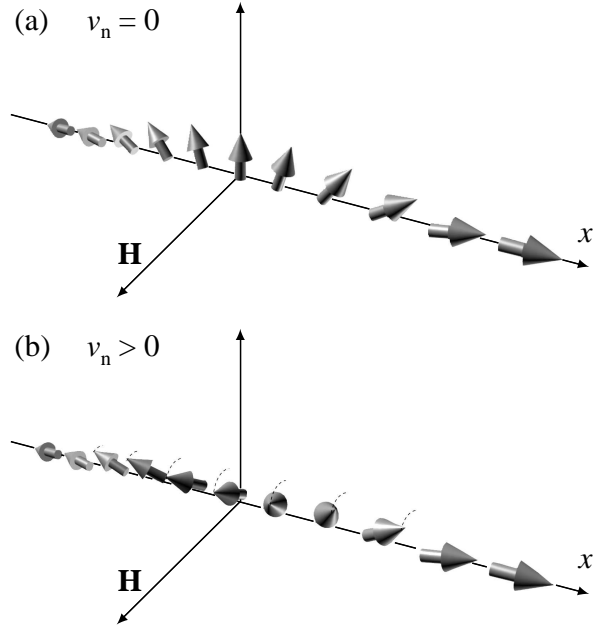}
\bigskip
\caption[composite]{ Distribution of the vector ${\hat{\bf l}}$ in a
one-dimensional soliton structure for $v_{\rm n}= 0$ (a) and for
$v_{\rm n}
\neq 0$ in a transverse magnetic field ${\bf H}$ (b).  The flow tends
to detach the texture from the plane $\perp {\bf H}$.
}\label{composite}\end{center}\end{figure}

In the previous section we studied the case where a flow was applied
to an initially uniform texture. Here we investigate some
cases where the initial state is inhomogeneous.  Let us consider a
texture where $\hat{\bf l}$ changes from the direction $-\hat{\bf x}$
to $\hat{\bf x}$, as depicted in Fig.\ \ref{composite}. 
Such a texture has the property, which follows directly from the
definition (\ref{e.phase}), that if it is rotated around
$\hat{\bf x}$ by angle
$\theta$, the phase difference
$\Delta\Phi$ changes by $2\theta$. Thus if nothing
prevents the rotation of the texture, the critical current vanishes
and the supercurrent is always dissipative. The presence of magnetic
field perpendicular to $\hat{\bf x}$ prefers to have $\hat{\bf d}$,
and via the dipole-dipole energy (\ref{fd}) also $\hat{\bf l}$, in the
plane perpendicular to ${\bf H}$. This gives rise to a finite critical
velocity that we aim to calculate. 

\begin{figure}[tbh]
\begin{center}\leavevmode
\includegraphics[width=0.8\linewidth]{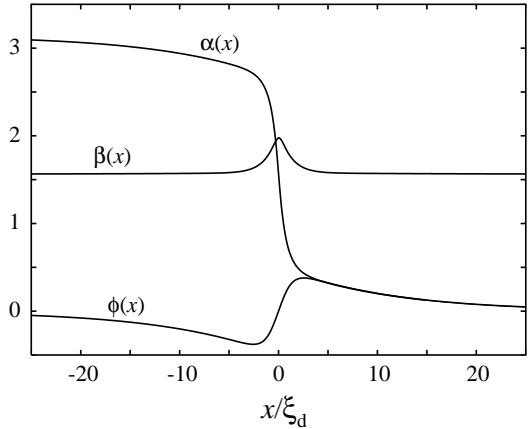}
\bigskip
\caption[kaikki]{ The structure of the unlocked soliton at $T=T_{\rm
c}$, $H \gg H_{\rm d}$ and 
$v_{\rm n}=0.15 v_{\rm d}$.   The angles $\alpha$, $\beta$, and $\phi$
are defined in Eqs.\  (\ref{lpara})-(\ref{dpara}) and $\psi=\pi/2$
because of the high-field limit. The fast change within $\vert
x\vert\lesssim\xi_{\rm d}$ is caused by the dipole unlocking in the
soliton, and the slower change outside is caused by the anisotropy of
the flow energy favoring
$\hat{\bf l}=\pm\hat{\bf x}$. }\label{unlocked}\end{center}\end{figure}

We study two different inhomogeneous structures. The first is known as
(dipole-unlocked) soliton \cite{MK78}. This is a domain-wall like
object where on one side
$\hat{\bf d}=\hat{\bf l}$ and on the other side $\hat{\bf d}=-\hat{\bf
l}$. Because the change between these two orientations costs
dipole-dipole energy, the thickness of the wall is on the order of the
dipole length
$\xi_{\rm d}$. In the absence of flow the asymptotic directions of
$\hat{\bf l}$ deviate from the normal $\pm\hat{\bf x}$ of the wall by
an angle that depends on the temperature. When a small flow is applied
perpendicular to the wall, the anisotropy of the kinetic energy
$\propto -(\hat{\bf l}\cdot{\bf v}_{\rm s})^2$ forces  the asymptotic
directions of $\hat{\bf l}$ to $\pm\hat{\bf x}$. The calculated
structure of the soliton is presented in Fig.\
\ref{unlocked}.  

\begin{figure}[tbh]
\begin{center}\leavevmode
\includegraphics[width=0.8\linewidth]{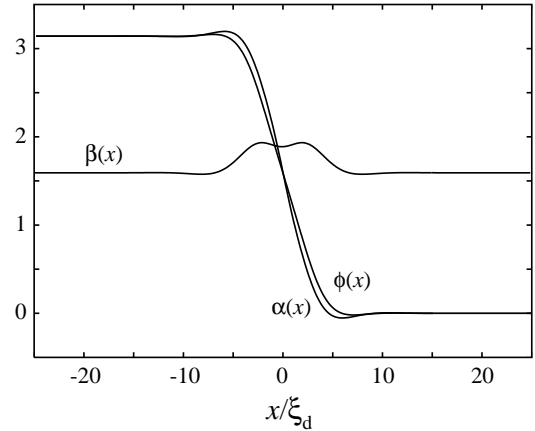}
\bigskip
\caption[kaikki]{  The structure of the locked soliton at $T=T_{\rm
c}$, $H \gg H_{\rm d}$ and 
$v_{\rm n}=0.8 v_{\rm d}$. The angles $\alpha$, $\beta$, and $\phi$
are defined in Eqs.\  (\ref{lpara})-(\ref{dpara}) and $\psi=\pi/2$.
The length scale is purely determined by the flow velocity, which here
is much larger than in Fig.\
\ref{unlocked}. }\label{locked}\end{center}\end{figure}

The second structure we study could be called a dipole-locked soliton.
The dipole-locking means that ${\hat{\bf d}}(x)\approx{\hat{\bf
l}}(x)$ everywhere. The region where
$\hat{\bf l}$ varies has finite length only in the presence of
flow.  The calculated structure of the locked soliton is presented in
Fig.\
\ref{locked}.  

\begin{figure}[tbh]
\begin{center}\leavevmode
\includegraphics[width=0.9\linewidth]{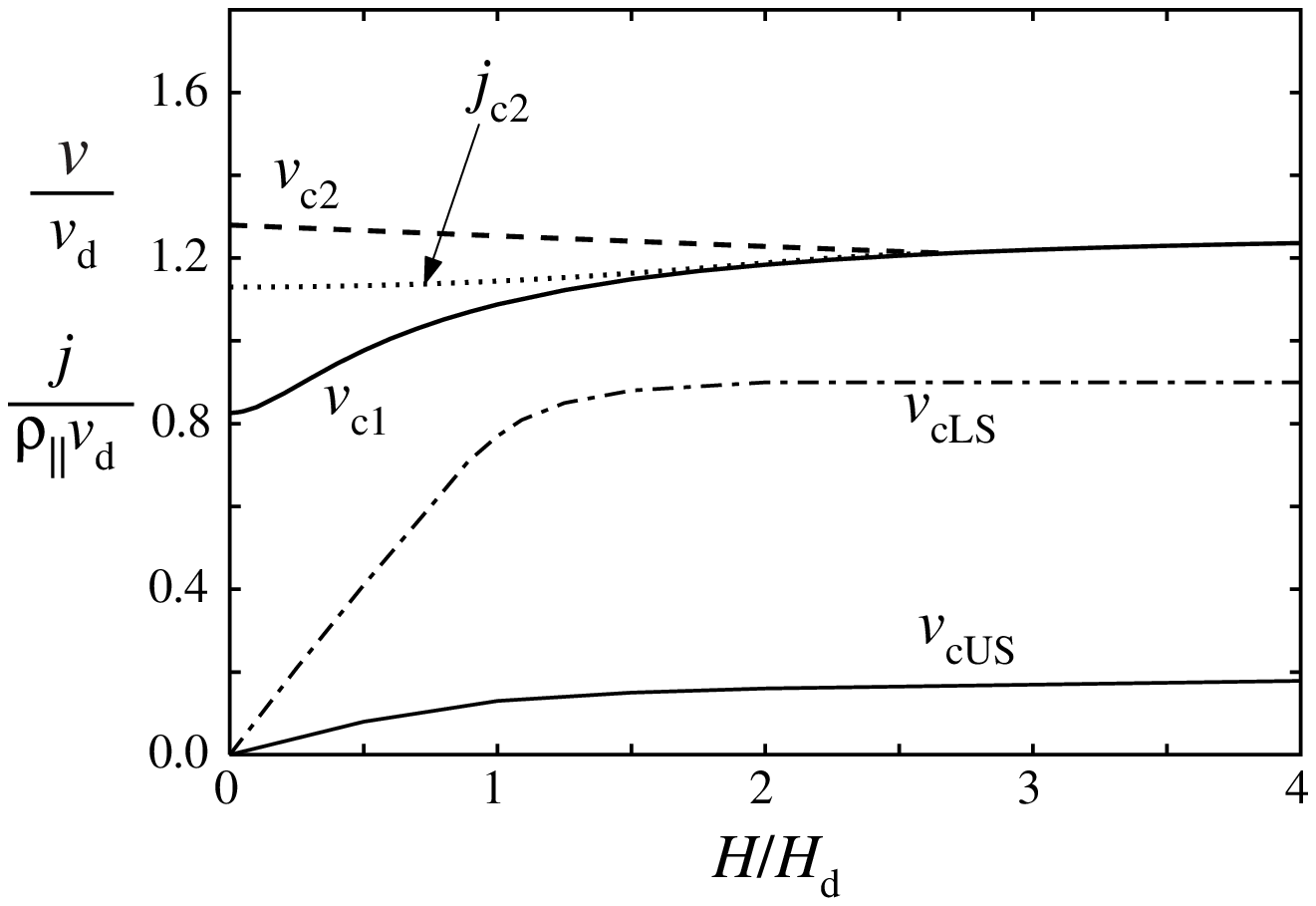}
\bigskip
\caption[kentat]{ The critical velocities of the locked soliton $v_{\rm
cLS}$, and the unlocked soliton $v_{\rm cUS}$ as a function of the
magnetic field $H$ at temperature
$T\approx T_{\rm c}$.  For comparison, we show also
$v_{\rm c1}$, $v_{\rm c2}$ and $j_{\rm c2}/\rho_\parallel$ (Fig.\
\ref{risto}). }\label{kentat}\end{center}\end{figure}

For both soliton structures the flow makes $\beta$ to deviate from
$\pi/2$. When the flow is further increased,
the structures become unstable against unlimited winding around the
flow direction. The critical values of $v_{\rm n}$ are denoted by
$v_{\rm cLS}$ for locked soliton and
$v_{\rm cUS}$ for unlocked soliton. The critical velocities  are
plotted in Figs.\ \ref{helical} and \ref{kentat}. 
The unlocked case has
previously been studied by Vollhardt and Maki \cite{Maki} at
$T\approx T_{\rm c}$ using a variational approach. Our calculations give
a much lower critical velocity than theirs. We note that the
same process that leads to $v_{\rm cUS}$ also determines the critical
velocity of a vortex sheet \cite{Sheet,VSexp}. 
The locked soliton has previously been studied only for $H=0$ or 
${\bf H}\parallel{\bf v}_{\rm s}$, where the critical velocity vanishes
and only dissipative state exists
\cite{HSol,DH,DL,1988}.

In the simulations the length $L$ of the computational region  has to
be chosen large in comparison to $\xi_{\rm d}v_{\rm d}/v_{\rm n}$. 
The fast variation in the unlocked soliton sets an upper limit for the
discretization length, which was typically chosen as
$0.1 \xi_{\rm d}$. If $L$ is not very long, the correct procedure is
to extract the critical current $j_{\rm cLS}$  (and $j_{\rm cUS}$)
from the numerical calculation and then find the critical velocity
using $v_{\rm cLS}=j_{\rm cLS}/\rho_\parallel$ (and $v_{\rm
cUS}=j_{\rm cUS}/\rho_\parallel$).

\section{Conclusions}
\label{conc}

We have studied different 1D textures and determined their stability
against 1D perturbations. It is reasonable to ask if limiting to 1D
perturbations is sufficient to determine the local stability. Namely,
we know at least one situation in $^3$He-A where this is not the case:
uniform $\hat{\bf l}\perp{\bf H} \parallel \hat{\bf x}$. Here the 1D
model above gives stability until $v_{\rm c1}=v_{\rm c2}=\sqrt{g_{\rm
d}/(\rho_\perp-\rho_\parallel)}$ (for $H\gg H_{\rm d}$), but allowing
${\bf v}_{\rm s}$ to deviate from the direction of the applied phase
difference gives instability at $v_{\rm n}$ that is by factor
$\sqrt{\rho_\parallel/\rho_\perp}$ lower \cite{Fetter2}. This situation
differs, however, from the ones studied in the previous sections. For
a homogeneous texture with ${\bf H} \perp
\hat{\bf x}$ there exists a strict proof that only 1D perturbations
are relevant \cite{LinLiu}. For helical and soliton textures the 1D
model allows a natural decay mechanism for the current, and we are not
aware of any mechanism that could give a lower critical velocity.
Therefore we believe that other than 1D perturbations are unimportant
for the helical and soliton textures studied above, although a strict
proof remains open. 

Measurements of the critical velocity are done by Ruu\-tu et al
\cite{Ruutu}. They study 
$^{3}$He-A in a circular cylinder that is rotated around its axis. In
the vortex-free container the relevant driving velocity $v_{\rm
n}=\Omega R$, where $\Omega$ is the angular velocity and $R$ the
radius of the cylinder. Because $R\gg\xi_{\rm d}$, the flow near the
cylindrical wall is
one-dimensional to a good approximation. The magnetic field along the
axis of the cylinder corresponds to   transverse field relative to the
flow along the whole perimeter of the cylinder, and thus the
calculations presented above should apply to this case. If the field is
perpendicular to the axis, all possible angles exist between the field
and flow. In this case the critical velocity of the ``uniform'' texture
is determined by the orientation
${\bf H}\parallel{\bf v}_{\rm s}$ \cite{Fetter2}, which gives a lower
value of $v_{c2}$ than ${\bf H}\perp{\bf v}_{\rm s}$.

Ruutu et al find a considerable spread in the critical velocities. On
one hand, our largest calculated values of $v_{\rm c}$ correspond to
the instability of the helical texture and coincide relatively well
with the largest values observed in the experiments. On the other
hand, the lowest measured critical velocities can be explained by
assuming the presence of a dipole-unlocked soliton.   Quantitative
comparison is given by Ruutu et al
\cite{Ruutu}.  The comparison supports the
basic assumption that the critical velocity in superfluid $^{3}$He-A
indicates an instability of the bulk, and it depends on the underlying
texture.  The bulk critical velocity is
quantitatively better understood in superfluid
$^{3}$He-A than in any other superfluid.

\section*{Acknowledgments}

This research was supported by Vilho, Yrj\"{o} and Kalle
V\"{a}is\"{a}l\"{a} Foundation and by the Academy of Finland.

\end{document}